# A rain induced landslide 3D model based on molecular dynamics with fractal and fractional water diffusion

Gianluca Martelloni[a*] , Franco Bagnoli[b] , Alessio Guarino[c]

[a]Department of Physics and Astronomy & CSDC (Center for the Study of Complex Dynamics), Via Sansone, 1 - 50019 Sesto Fiorentino (FI), Italy.
gianluca.martelloni@unifi.it

[b]Department of Physics and Astronomy & CSDC (Center for the Study of Complex Dynamics), Via Sansone, 1 - 50019 Sesto Fiorentino (FI), Italy.
franco.bagnoli@unifi.it

[c]ICARE Laboratory, University of Reunion Island, 1 allée des aigues marines, 97400 Saint Denis de la Réunion, Reunion Island, France.
alessio.guarino@univ-reunion.fr

**Abstract**

We present a three-dimensional model, based on cohesive spherical particles, of rain-induced landslides. The rainwater infiltration into the soil follow the either the fractional or the fractal diffusion equations. We solve analytically the fractal diffusion partial differential equation (PDE) with particular boundary conditions to simulate a rainfall event. Then, for the PDE, we developed a numerical integration scheme that we integrate with MD (Molecular Dynamics) algorithm for the triggering and propagation of the simulated landslide. Therefore we test the numerical integration scheme of fractal diffusion equation with the analytical solution. We adopt the fractal diffusion equation in term of gravimetric water content that we use as input of triggering scheme based on Mohr-Coulomb limit-equilibrium criterion, adapted to particle level. Moreover, taking into account an interacting force Lennard-Jones inspired, we use a standard MD algorithm to update particle positions and velocities. Then we present results for homogeneous and heterogeneous systems (i.e. composed by particles with same or different radius respectively). Interestingly, in the heterogeneous case, we observe segregation effects due to the different volume of the particles. Finally we show the parameter sensibility analysis both for triggering and propagation phase. Our simulations confirm the results of our previous two-dimensional model and therefore the feasible applicability to real cases.

**Keywords :**

Fractional and fractal PDE

Molecular dynamics

Landslides

Computational model

---

[*] Corresponding author :  gianluca.martelloni@unifi.it

# 1 Introduction

Landslides are extreme events [1,2] that often cause environmental damages and human losses. In Sarno, Italy, a single landslide caused 160 victims in 1998, [3]. Landslide prevention is a problem which has received a lot of attention in a variety of fields: civil protection [4,5], urban planning [6] and scientific research [7-9]. The scientific research usually focus on the understanding of the complex phenomenology observed in a landslide [10], on the reproduction of landslide formation and dynamics [11-13] and on the prediction of a landslide trigger [14-16]. Despite great experimental, theoretical and numerical efforts [17], many aspects still remain unclear about the landslide process itself.
A major cause of the landslides triggering is represented by hydrogeological factors (surface and groundwater flow, amount and distribution of internal pressure) which are caused by rainfall. The latter point has lately received a lot of attention of the scientific community. Some empirical models [18-23] allow to define, on a statistical basis, rainfall thresholds for the prediction of landslides events. These thresholds are used by authorities for population protection. Raising computing power has permitted increasingly sophisticated numerical simulations to predict landslides dynamics and trigger. Most of landslides trigger prediction models are based on Richards equation for water infiltration [24,25]. The models used to study landslide dynamics are built on either a fluid-like eulerian approach, [11,26] or a granular-lagrangian one [27-30].
The main motivation of this work is to present a three-dimensional model for rainfall triggered landslides based on cohesive spherical particles that, initially, are at equilibrium.

# 2 Materials and method

In previous works [29,30], we proposed a 2D rain induced landslide model for shallow landslides (without water diffusion, but with lubrication effects) and one model for "deep" landslide (considering infiltration effects modeled by means of classical diffusion equation, see [25]). Here, we present a 3D molecular dynamics model with rainwater anomalous diffusion schematized by fractal diffusion equations [31]. The model is mesoscopic and consists of a lagrangian schematization of the soil based on cohesive spherical interacting particles, initially placed in equilibrium (see an example of initially configuration in figure 1). The output of the fractal infiltration scheme represents the input of the triggering model based on Mohr-Coulomb criterion applied at particle level [30,32], while the particle positions are updated using a first and second order algorithm.

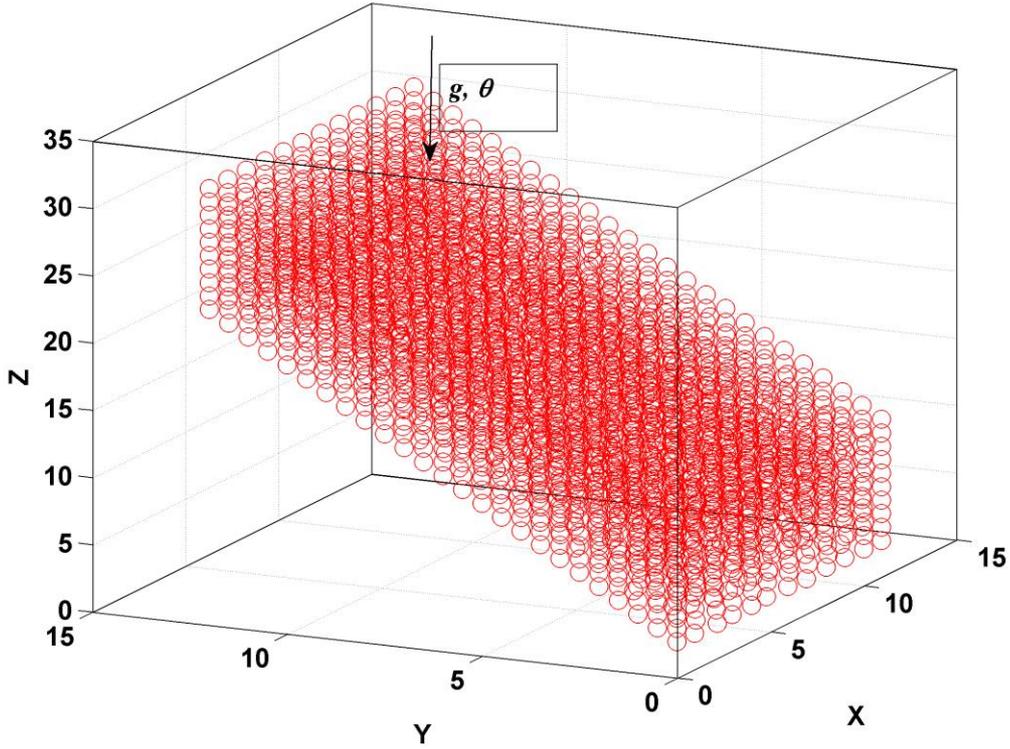

**Fig. 1**: example of initially configuration of the system with all particles at rest placed on an inclined plane with vertical infiltration due to the simulated rainfall. In this figure the particle radii are not explicated as we use a simple representation with circles.

**2.1 Hydrological model**

We simulate the effects of infiltration into the soil, due to rainfall, by means of the fractal and fractional diffusion equations which have found wide application in recent decades in the study of these processes starting from physical justifications, i.e., the anomalous diffusion [33]. The porous materials, as the soils, have a natural fractal structure, i.e., have non-integer dimension that gives a measure how the space is filled by the material itself. Moreover, the diffusion equations derived in fractals and fractional form are mathematically compatible with no-Gaussian statistic of diffusion processes where the mean square displacements of the particles fluid trajectory exhibit a power lows time dependence. While the classical diffusion equation is compatible with the central limit theorem for which we have a Gaussian distribution of spatial jumps with linear time dependence of mean square displacements [34]. Therefore, in general, anomalous diffusion (sub or super-diffusion) deal with the break of the central limit theorem and the corresponding processes do not follow a Gaussian statistical type [35,36]. Furthermore experimental evidence show the correctness of a fractal-fractional approach to porous systems [37] and specifically for the infiltration in the unsaturated soil [31,38]. Thus, fractional and fractal diffusion equations, applicable to volumetric (or gravimetric) water content $\theta$ as shown in [31], can be expressed respectively by the following equations, where we neglect the term of gravity [25] and we consider only vertical infiltration (one spatial dimension),

$$\frac{\partial^{\alpha_1}}{\partial t^{\alpha_1}}\theta(z,t) = D_1 \frac{\partial^{\beta_1}\theta(z,t)}{\partial z^{\beta_1}}, \quad \frac{\partial}{\partial t^{\alpha_2}}\theta(z,t) = D_2 \frac{\partial}{\partial z^{\beta_2/2}}\left(\frac{\partial}{\partial z^{\beta_2/2}}\theta(z,t)\right). \quad (1)$$

In equations (1), *t* represents the time, *z* the depth, ($\alpha_1,\beta_1$) and ($\alpha_2,\beta_2$) are not integer (for ($\alpha_1,\beta_1$)=(1,2) and ($\alpha_2,\beta_2$)=(1,2) we obtain the classical diffusion equation), while $D_1$ ($L^{\alpha 1}T^{-\beta 1}$) is the fractional diffusion coefficient in the first of eq.(1) and $D_2$ ($L^{\alpha 2}T^{-\beta 2}$) the fractal diffusion coefficient in the second one. Considering the fractional equation we have sub-diffusion if $2\alpha_1<\beta_1$, super-diffusion if $2\alpha_1>\beta_1$, while in the fractal we have sub-diffusion if $\alpha_2 < 2\beta_2/(6-\beta_2)$ and super-diffusion if $\alpha_2 > 2\beta_2/(6-\beta_2)$ as shown in [34]. The fractional time derivative describes memory effects, the fractional Laplacian the non-locality of the phenomenon, while the fractal time derivative and the Laplacian describe respectively the magnitude variations with respect to temporal and spatial ones whose exponents $\alpha_2$ and $\beta_2$ are related to the fractal dimension of the system, as well as $\alpha_1$ and $\beta_1$ of the fractional equation [39]. Furthermore we note that there are equivalences between classes of fractional and fractal equations in terms of the fundamental solution (stretched Gaussian), as for example in [40].

The fractional diffusion equation, with the diffusion coefficient constant, can be solved analytically for different boundary conditions as shown in [41]. The fractal one, denoting with *D* the fractal diffusion coefficient and with $\alpha$ and $\beta$ the time and spatial derivative order respectively, can be solved analytically with a simple change of variables ($t^* = t^\alpha$, $z^* = z^{\beta/2}$) as shown in [34]. Moreover with another change of variables we can find the analytical solution considering *D* non constant [42]. In general *D* depends directly on the water content [31], but here, for simplicity and in the absence of experimental data, we consider *D* constant or depth dependent [43]. In the case of fractal equation we develop a numerical integration method (implicit scheme of Adams-Bashforth-Moulton), both to integrate the infiltration scheme in the calculation of the landslide triggering discrete model and to consider a stochastic forcing $\delta(z,t)$, required for modeling the heterogeneity of a porous system, i.e. a soil, and also to consider different boundary condition. Therefore the second of eq. (1) becomes:

$$\frac{\partial}{\partial t^\alpha}\theta(z,t) = D(z)\frac{\partial}{\partial z^{\beta/2}}\left(\frac{\partial}{\partial z^{\beta/2}}\theta(z,t)\right) + \delta(z,t). \tag{2}$$

We note that the eq. (2) is analogous to the fractional one as shown in [43]. To test the numerical integration scheme we solve analytically the second of eq. (1) by means of Fourier series method considering *D* constant, $\delta(z,t) = 0$ and $\beta=2$. In particular we use non homogeneous boundary conditions ($\theta(0,t)=a\cdot t$, $\theta(L,t)=d$, $\theta(z, 0)=0$). The first boundary condition represents the schematization of progressive soil saturation (rain induced) starting from ground-level, the rain is assumed uniform (*a* = constant); the second one is the condition at depth *L* where the soil is saturated due to the presence of a groundwater (*d* constant $\neq$ 0), or with a rock layer (*d* = 0); finally the third one is the initial condition of dry soil. Therefore we solve the following problem:

$$\begin{cases}\frac{\partial}{\partial t^\alpha}\theta(z,t) = D\frac{\partial^2}{\partial z^2}\theta(z,t) \\ \theta(0,t) = a\cdot t \quad \theta(z,0) = 0 \quad \theta(L,t) = d\end{cases}. \tag{3}$$

Consequently applying the variable transformation $t^* = t^\alpha$, we obtain the new problem:

$$\begin{cases}\frac{\partial}{\partial t^*}\theta(z,t^*) = D\frac{\partial^2}{\partial z^2}\theta(z,t^*) \\ \theta(0,t^*) = a\cdot(t^*)^{1/\alpha} \quad \theta(z,0) = 0 \quad \theta(L,t^*) = d\end{cases}. \tag{4}$$

Now it is possible to apply, at the latter problem, the method of Fourier series to obtain the following solution replacing $t^*$ with $t^\alpha$:

$$\theta(z,t) = a \cdot t\left(1 - \frac{z}{L}\right) + d \cdot \frac{z}{L} + \sum_{n=1}^{\infty} \left[ \begin{array}{l} (-1)^{1-1/\alpha} C \cdot \Gamma\left(\frac{1}{\alpha}, -B\right) \cdot B^{-1/\alpha} + \\ -\left((-1)^{1-1/\alpha} C \cdot \Gamma(1/\alpha) \cdot B^{-1/\alpha} + \frac{2d \cdot \cos(\pi \cdot n)}{\pi \cdot n}\right) \end{array} \right] \cdots$$

$$\cdots \exp(-B \cdot t^\alpha) \cdot \sin\left(\frac{n \cdot \pi \cdot z}{L}\right) \quad , \quad (5)$$

$$B = D \cdot \left(\frac{\pi \cdot n}{L}\right)^2, \quad C = -\frac{2a}{n \cdot \pi \cdot \alpha}, \quad \Gamma(s,x) = \int_x^\infty \exp(-t) \cdot t^{s-1} dt$$

where $\Gamma(s,x)$ is the incomplete gamma function and $\Gamma(1/\alpha)$ is the gamma function calculated in $1/\alpha$. In the Figure 2 we compare the theoretical solutions and numerical one and some results are reported for the normal and sub-diffusive case: as one might expect, in the latter, $\theta$ is minor than the classical one. Then in Figure 3 we show some other cases obtained by means of numerical integration: in particular we consider fractal laplacian in eq. 2 ($\alpha=1$, $\beta\neq 2$) and we analyze the response of eq.(2) considering also $\delta(z,t)\neq 0$ and $D=D_0 \cdot z^{-\gamma}$ ($D_0$ and $\gamma$ are constants), i.e., the diffusion coefficient varies according to a power law, and finally an example of super-diffusive case with $\beta=1.6$.

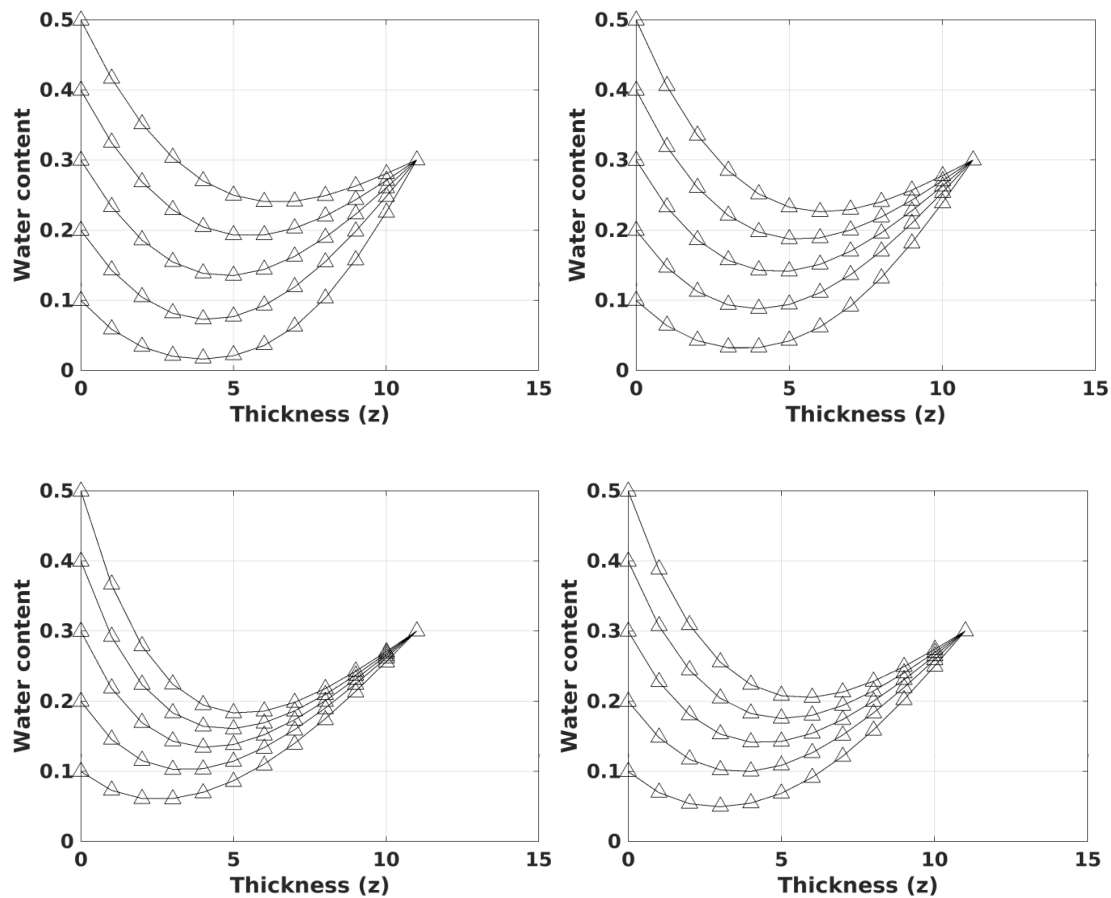

**Fig. 2**: comparison between theoretical (solid line) and numerical solution (upward-pointing triangle) of fractal diffusion equation with $\beta=2$, $D$=constant and $\delta(z,t)=0$ – from the top left clockwise we have respectively $\alpha=1$ (normal diffusion), $\alpha=3/4$, $\alpha=1/2$ and $\alpha=1/3$ (sub-diffusion in the last three cases).

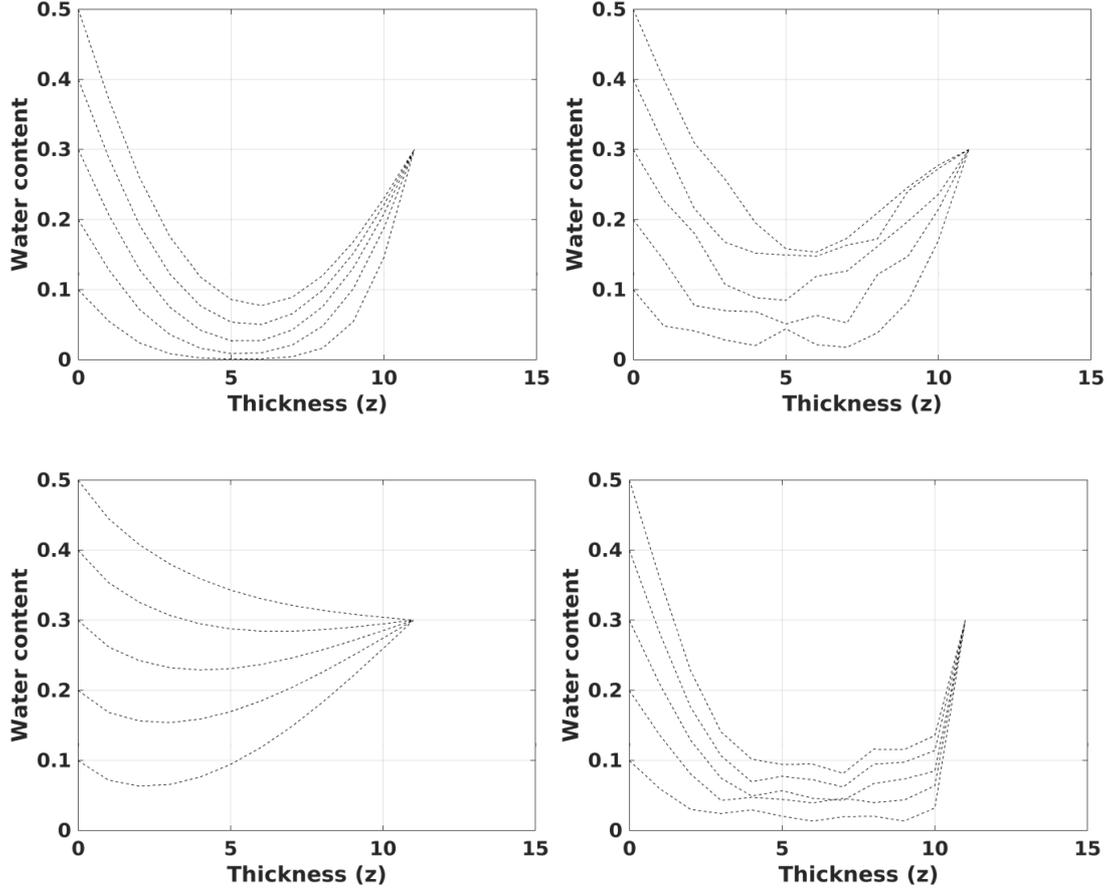

**Fig. 3**: numerical solution of fractal diffusion equation with $\alpha$=1 – from the top left clockwise we have respectively $\beta$=2.5 ($D$=constant and $\delta(z,t) = 0$, sub-diffusive case), $\beta$=2.5 ($D$=constant and $\delta(z,t)\neq 0$, sub-diffusive case), $\beta$=2.5 ($D=D_0 \cdot z^{-\gamma}$ (where $D_0$ and $\gamma$ are constants) and $\delta(z,t)\neq 0$, sub-diffusive case) and $\beta$=1.6 ($D$=constant and $\delta(z,t) = 0$, super-diffusive case).

**2.3 Triggering and propagation molecular dynamics scheme**

The triggering scheme is based on the Mohr-Coulomb law [30],
$$\tau_f = (\sigma - p) \cdot \tan\phi' + c' \qquad (6)$$
where $\tau_f$ is the shear stress at failure, $\sigma$ the normal stress, $\phi'$ the friction angle and $c'$ the cohesion term. As the eq. (6) represents a simple friction law, short of the term of cohesion, it can be adapted for a single particle, rewriting eq. (6) as
$$\begin{cases} \tau_{f,i} = F_{s,i} + c' \\ F_{s,i} = \left( M_i(x,y,z,t) \cdot g \cdot \cos(\varphi) - p(x,y,z,t) \right) \cdot \tan\phi' \end{cases} \qquad (7)$$
In eq. (7) $\varphi$ is the slope angle, $M_i$ represents the total mass $m_i + w_i(t)$, i.e., respectively the dry mass of each particle and the cumulative amount of absorbed water per particle ($w_i = \theta \cdot m_i$), while $p(x,y,z,t)$ is the pore pressure interpreted as a scalar field (see [30]).

The cohesion term is not constant in our simulations, as it is affected by water content: in particular cohesion decreases quickly with $\theta$ as show in [44] and therefore, in absence of experimental data, we consider $c'$ exponentially dependent on $\theta$, i.e., $c' = c_k \cdot exp(-k_c \cdot \theta)$, where $k_c$ is constant, while $c_k$ varies linearly with depth to take into account the higher consolidation of the deeper soil layers, therefore higher cohesion with depth. This choice is due to linear relation between cohesion and critical shear stress as shown in [45].

To explicit the pore pressure $p$ as a function of the gravimetric water content $\theta$ (the ratio between water mass $w_i$ and dry mass $m_i$) we adopt the simple relation of hydrostatic law:

$$p(x,y,z,t) = k_p \cdot \theta(x,y,z,t) \cdot m_i(x,y,z) \cdot g \cdot z$$
$$\theta(x,y,z,t) = k(x,y) \cdot \theta(z,t)$$
(8)

where $k_p$ is a normalization constant (volume dimension) and $k(x,y)$ is a stochastic function generated by means a Gaussian random number generator in order to extend the water content in three dimension and considering the natural variability of the infiltration processes in the soil.

Therefore the triggering mechanism is defined for each particle $i$ [30,32] and based also on a velocity threshold $v_d$:

$$|\mathbf{F_i}| < F_{si} + C'$$
$$|\mathbf{v_i}| < v_d$$
(9)
$$\mathbf{F_i} = \mathbf{F_{gi}} + \sum_{j=1}^{j=n_k} \mathbf{F_{ij}}$$

where $\mathbf{F_i}$ represents the active force on particle $i$, i.e., the force of gravity $\mathbf{F_{gi}}$ plus a force $\mathbf{F_{ij}}$, similar to a Lennard-Jones potential, expressed by means of the following pseudo-code:

$$\begin{aligned}
&\text{if} \quad r_{ij} < D \\
&\mathbf{F_{ij}} = -k_1 \cdot \left(\frac{r_{ij}}{D}\right)^{-2} \hat{\mathbf{r}}_{ij} \\
&\text{else} \\
&\mathbf{F_{ij}} = k_2 \cdot \left(\frac{r_{ij}}{D}\right)^{-2} \hat{\mathbf{r}}_{ij} \\
&\text{end}
\end{aligned}$$
(10)

$$\mathbf{F_{ij}} = -\mathbf{F_{ji}}$$
$$|\hat{\mathbf{r}}_{ij}| = 1$$
$$r_{ij} = |\mathbf{r_{ij}}| = \sqrt{(x_j-x_i)^2 + (y_j-y_i)^2 + (z_j-z_i)^2}$$

In eq. (10) $\mathbf{F_{ij}}$ is the interaction force and it is interpreted as the force that acts on particle $i$ due to $j$ one, $r_{ij}$ is the distance between the two center of mass, $D$ is the diameter (constant=1) of the particles, while $k_1$ and $k_2$ are constants.

Moreover we utilize another expression for the interaction force varying the radius of the particles (small stochastic variation around 0.5), i.e., we consider this pseudo-code (see simulations and results section):

$$\begin{aligned}
&if \quad r_{ij} < R_i + R_j\\
&\mathbf{F}_{ij} = -k_1 \cdot \left(\frac{r_{ij}}{R_i + R_j}\right)^{-2} \hat{\mathbf{r}}_{ij}\\
&elseif \quad r_{ij} = R_i + R_j\\
&\quad \mathbf{F}_{ij} = 0\\
&else\\
&\mathbf{F}_{ij} = k_2 \cdot \left(\frac{r_{ij}}{R_i + R_j}\right)^{-2} \hat{\mathbf{r}}_{ij}\\
&end
\end{aligned} \qquad (11)$$

where $R_i$ and $R_j$ represent the radius of two interacting particle (see Figure 4). The latter expression is in agreement with the common technique used for the collisions in a granular matter [46] and it is equivalent to the classical cross section method for the diffusion of a particle in a material medium in which the spreading particle is considered point-like and its radius is added to each particle of the medium.

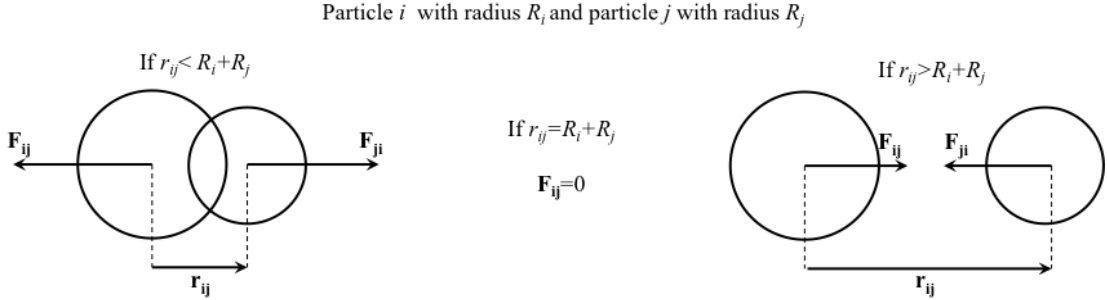

**Fig. 4**: geometric 2D description of interaction force acting between two spherical particles.

Then for the moving particles we take into account, in addition to the third of eq. (9), the dynamic friction and viscosity force, proportional to velocity by means of coefficient $\gamma$, i.e.,

$$\mathbf{F} = \mathbf{F_i} + \mathbf{F_{di}} + \mathbf{F_{ji}} = \mathbf{F_{gi}} + \sum_{j=1}^{j=n_k} \mathbf{F_{ij}} + \mathbf{F_{di}} - \gamma \cdot \mathbf{v_i}, \qquad (12)$$

where the dynamic friction force, acting on the particles in contact with the sliding surface, is expressed by

$$\begin{aligned}
\mathbf{F_{di}} &= M_i(t) g \cos(\alpha) \cdot \mu_D \cdot (-\hat{\mathbf{v}})\\
\mu_D &= (\mu_d \cdot \exp(-k_\theta \cdot \theta(t)) + \mu_{dlow} \cdot (1 - \exp(-k_\theta \cdot \theta(t))))
\end{aligned}, \qquad (13)$$

where $\mu_D$ is the friction dynamic coefficient, $\mu_d$ and $\mu_{dlow}$ are respectively the upper and lower limit of $\mu_D$. Moreover the eq. (13) is similar to the one used in [30], but here the friction dynamic force is set in relation with the gravimetric water content $\theta$.

The positions $\mathbf{r}$ and the velocities $\mathbf{v}$ of the masses are updated with a standard molecular dynamics scheme based respectively on algorithm of the first or second order (that take into account a better approximation in case of the dumping term presence, as in our case):

$$\begin{cases}
\mathbf{r}(t + \Delta t) = \mathbf{r}(t) + \mathbf{v}(t)\Delta t\\
\mathbf{v}(t + \Delta t) = \mathbf{v}(t) + \dfrac{1}{m}\mathbf{F}(t)\Delta t
\end{cases}. \qquad (14)$$

$$\begin{cases} \mathbf{a}(t) = \mathbf{F}(t)/m \\ \mathbf{r}(t+\Delta t) = \mathbf{r}(t) + \mathbf{v}(t)\Delta t + \frac{1}{2}(\mathbf{a}(t))\Delta t^2 \\ \hat{\mathbf{v}}(t+\Delta t) = \mathbf{v}(t) + \frac{1}{2}[\mathbf{a}(t) + \mathbf{a}(\mathbf{r}(t+\Delta t), \mathbf{v}(t)+\mathbf{a}(t)\Delta t, t+\Delta t)]\Delta t \\ \mathbf{v}(t+\Delta t) = \mathbf{v}(t) + \frac{1}{2}[\mathbf{a}(t) + \mathbf{a}(\mathbf{r}(t+\Delta t), \hat{\mathbf{v}}(t+\Delta t), t+\Delta t)]\Delta t \end{cases} \quad (15)$$

In eqs. (14-15) **v** is the velocity, **a** the acceleration and **F** the resulting force. Moreover we test a symplectic version of the algorithm described by eq. (15) and we observe very similar results due to the dissipative term of eq. (12).

Finally the particle positions are recalculated within a sphere of assigned radius $L_r$, according to the illustrative scheme of Figure 5:

$$(x-x_i)^2 + (y-y_i)^2 + (z-z_i)^2 \leq L_r^2, \quad L_r = \sqrt{2}. \quad (16)$$

In eq. (16) we consider $L_r$ equal to the square root of 2 according to the initial interaction to second neighbors in a regular grid.

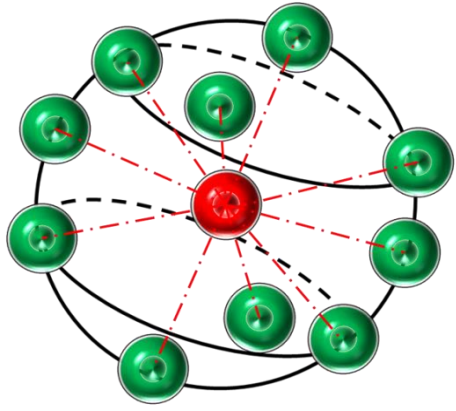

**Fig. 5**: scheme of interaction re-calculus for each particles of the system according to the action and reaction principle.

Finally, regarding the landslide propagation, i.e., after the triggering phase, it is possible to consider an effective one dimensional model of the movement along an inclined plane. Indeed, obviously, we do not consider the interaction force, that are internal one, and the second principle of dynamics is expressible by means of the gravity, the frictional and the viscosity force:

$$F = \langle m \rangle \cdot a = F_g + F_d - \gamma \cdot v = \langle m \rangle \cdot g \cdot \sin(\varphi) - \langle \mu_D \rangle \cdot \langle m \rangle \cdot g \cdot \cos(\varphi) - \gamma \cdot v, \quad (17)$$

where $\langle m \rangle$ is the average mass of the particles and $\langle \mu_D \rangle$ the average friction dynamic coefficient. Therefore we obtain the simple following one dimensional equation of motion for the velocity $v$ along the inclined plane:

$$\frac{dv}{dt} = g \cdot \sin(\varphi) - \langle \mu_D \rangle \cdot g \cdot \cos(\varphi) - \frac{\gamma}{\langle m \rangle} \cdot v. \quad (18)$$

The eq. (17) is a differential equation of first order, whose solution is given by

$$v = \left( v_0 - \frac{\langle m \rangle \cdot g \cdot (\sin(\varphi) - \langle \mu_D \rangle \cdot \cos(\varphi))}{\gamma} \right) \cdot \exp\left( -\frac{\gamma}{\langle m \rangle} \cdot t \right) + \\ + \frac{\langle m \rangle \cdot g \cdot (\sin(\varphi) - \langle \mu_D \rangle \cdot \cos(\varphi))}{\gamma}, \quad (19)$$

where $v_0$ is the initial condition, i.e., the velocity at the failure instant.

## 3 Simulations and results

In this section we show some numerical experiments and we present the main results, e.g., parametric sensibility of the model and statistical analysis. In particular we present the results for two landslide simulation respectively with the same and different radii of the spherical particles. As in the previous works regarding two-dimensional schemes [29,30], we imposed stochastic variation for the model parameters according to variability of real slopes. In particular the multiplicative factor of cohesion function $c_k$ varies with coordinates $x$ and $y$ of the reference system of the inclined plane according to $c_k = c_{k0} + 0.05 \cdot c_{k0} \cdot rn_1(x,y)$, where $rn_1$ is a uniform random number ($rn_1 \in [0,1]$). We use a uniform random number generator as $c_{k0}$ represents a static minimum threshold that regulates the triggering mechanism according to Mohr-Coulumb failure criterion. Then the dynamic parameter $\mu_d$ and $\mu_{dlow}$ vary respectively following the functions $\mu_d = \mu_{d0} + 0.1 \cdot \mu_{d0} \cdot rgn_1(t,x,y)$, $\mu_{dlow} = \mu_{dlow0} + 0.1 \cdot \mu_{dlow0} \cdot rgn_2(t,x,y)$, where $rgn_1(t,x,y)$ and $rgn_2(t,x,y)$ are Gaussian random number depending on time and coordinates $x$ and $y$. Also the particle positions, as we can see in figure 1, are initialized casually, with variations (normally distributed) in a range of 20% compared to a regular grid. The particle masses varies depending on numerical experiments, while the other parameters (slope, friction angle, fractal order derivatives, fractal diffusion coefficient, viscosity coefficient, interacting force coefficients and gravity acceleration) are constant in a single simulation.

### 3.1 Simulation 1

In this simulation we consider a system of spherical particles with the same radius in a random configuration, arranged in an inclined plane with vertical infiltration, as shown in Figure 6. Moreover we impose rigid boundary condition, i.e., the particles are confined within an open "container" so that the particles can slide along the inclined plane. In this simulation we use the force scheme of eq. (10) and the first order algorithm to update positions and velocities (eq. 14) with temporal step $\Delta t = 0.001$. We simulate a system of 3000 particles with a duration of $10^5$ temporal steps, i.e., $10^2$ in time). The masses are varied in a range of 40% (normally distributed) respect to a reference value of the particle mass $m_0$ (with $m_0 = 0.007$ the masses vary in the range ~[0.0048, 0.0097]). As the model is theoretical and not applied to real case at the moment, we do not express the unit of measurement of the physical magnitudes involved in the model. Therefore we adopt the values indicated in Table 1.

| | |
|---|---|
| Slope angle ($\varphi$) | 60° |
| Friction angle ($\phi'$) | 40° |
| Fractal order temporal derivative ($\alpha$) | 1 |
| Fractal order space derivative ($\beta$) | 2.5 |
| Fractal diffusion coefficient ($D$) | 100 |
| Repulsive coefficient of interacting force ($k_1$) | 0.006 |
| Attractive coefficient of interacting force ($k_2$) | 0.002 |
| Gravity acceleration ($g$) | 1 |
| Coefficient of cohesion ($c_{k0}$) | 0.5 |
| Upper friction coefficient ($\mu_{d0}$) | 0.69 |
| Lower friction coefficient ($\mu_{dlow0}$) | 0.4 |
| Viscosity coefficient ($\gamma$) | 0.01 |

**Table 1**: values of the model constant for simulation 1.

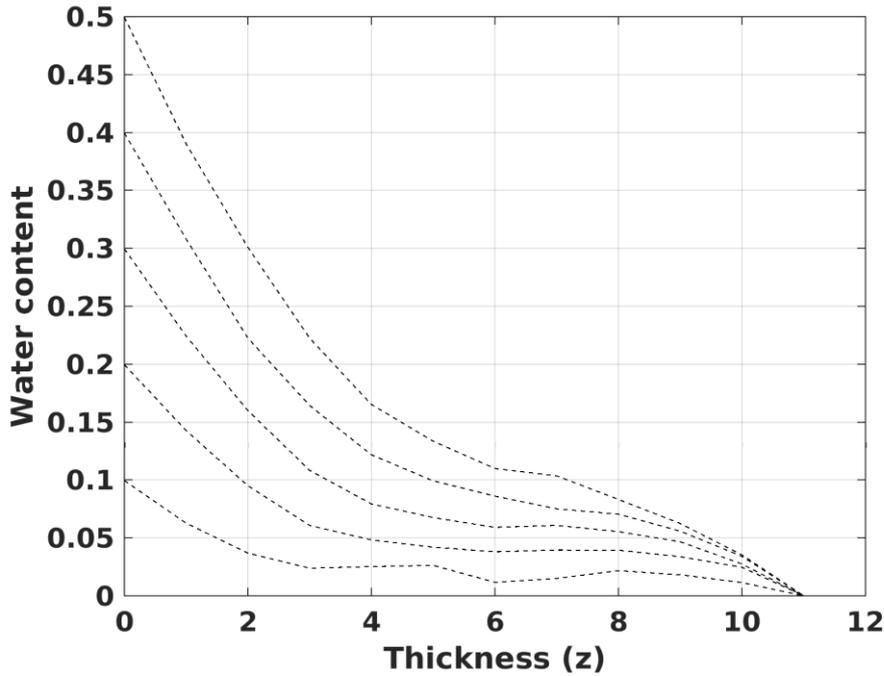

**Fig. 6**: water content in space (thickness of the system) and in time (curve from bottom to top of the figure) for a single vertical layer in simulation 1.

In Figure 6 we show the trend of water content curve in time and space for a single vertical layer. We adopt non homogeneous boundary conditions ($\theta(0,t)=a \cdot t$, $\theta(L,t)=d$, $\theta(z, 0)=0$) with $d = 0$ as explain above. The duration of the rain is imposed to $10^3$ time steps in order to induce the triggering, according to the choice of the model parameters. In Figure 7 we report some configuration of the system at different times, where the red particles are at rest and the green ones are in motion. Moreover the right bottom graph of Figure 7 represents the final configuration (the same of bottom left graph) exploiting the real dimension of the slope and masses of the landslide. In the latter figure we note the progressive infiltration of water that perturbs the pore pressure field causing the motion of the particles. Once the landslide go out from rigid boundary, it expands with height reduction as it occurs in real cases or in setup experiments [26]. As in [30], we measure the distribution of the time intervals between two subsequent triggering events of particle motion starting. A power law distribution of these durations is observed in this simulation (see Figure 8). Then we analyze the behavior of mean energy in correspondence of the time of subsequent particle triggering. We report in Figure 9 the mean kinetic energy and the mean kinetic energy increments, observing a rapid increasing due to landslide triggering. In Figure 10 we report the distribution of mean kinetic increments during the triggering phase, observing a transition in time from Gaussian to power law. This result is consistent with our simulations for two-dimensional landslide models [30]. Finally in Figure 11 we show the comparison between the mean velocity (component along inclined planed) obtained from this simulation (green line) and the velocity of the one-dimensional effective model. The result is similar and demonstrate the consistent of our simulation. Obviously the effective model cannot consider all 3D effects. We also analyze the distribution of each component of velocity and possible transition in time (we do not report the correspondent trend). The component $v_x$ and $v_y$ present Gaussian distribution, only the component $v_z$ exhibit a transition in time from Gaussian to power law due to repulsive force, i.e., we observe many small velocities along the $z$ axis of the inclined plane.

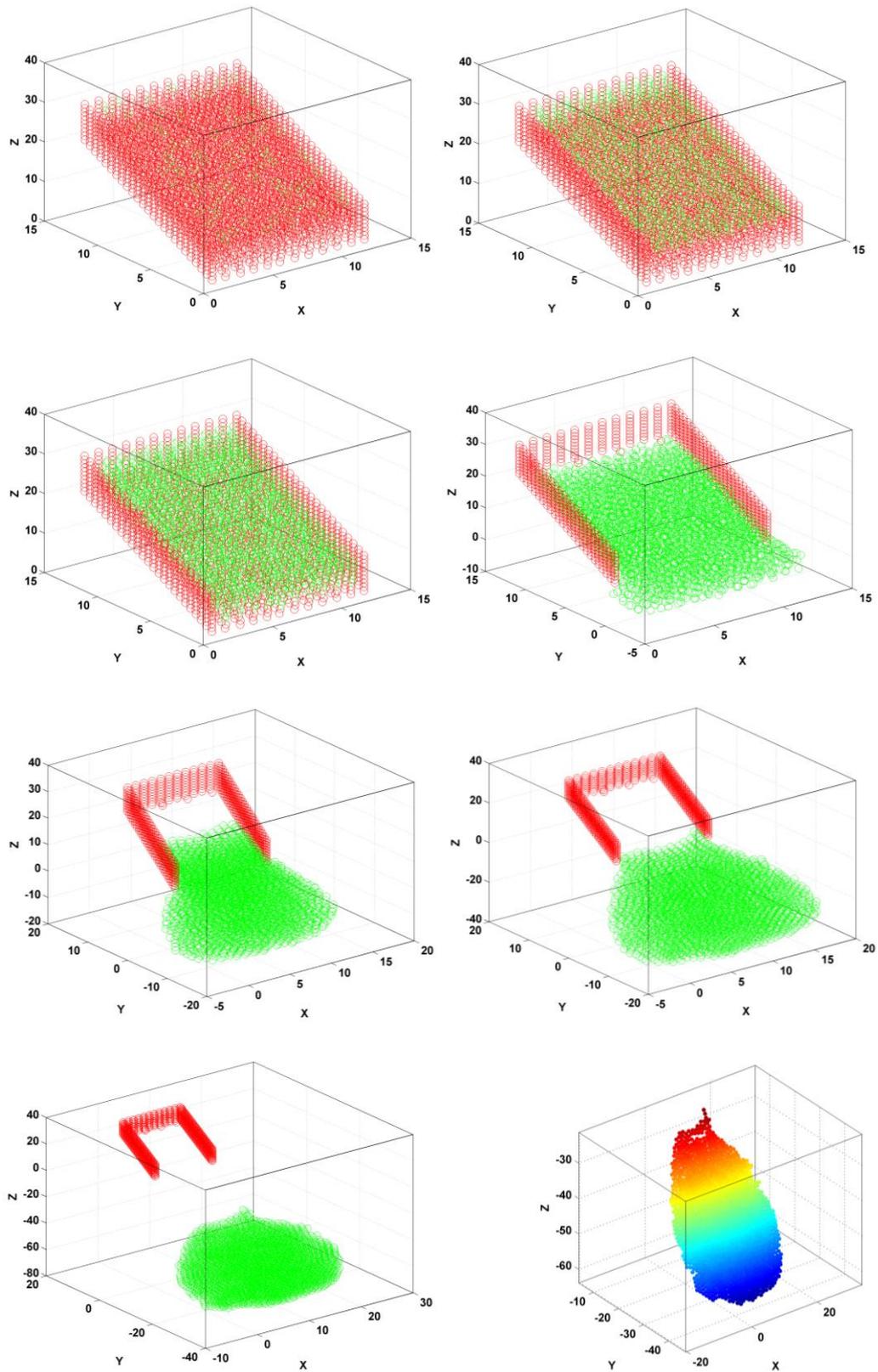

**Fig. 7**: configuration of the landslide at different time starting from left to right and from top to bottom. Red particles are at rest, while green particles are in motion.

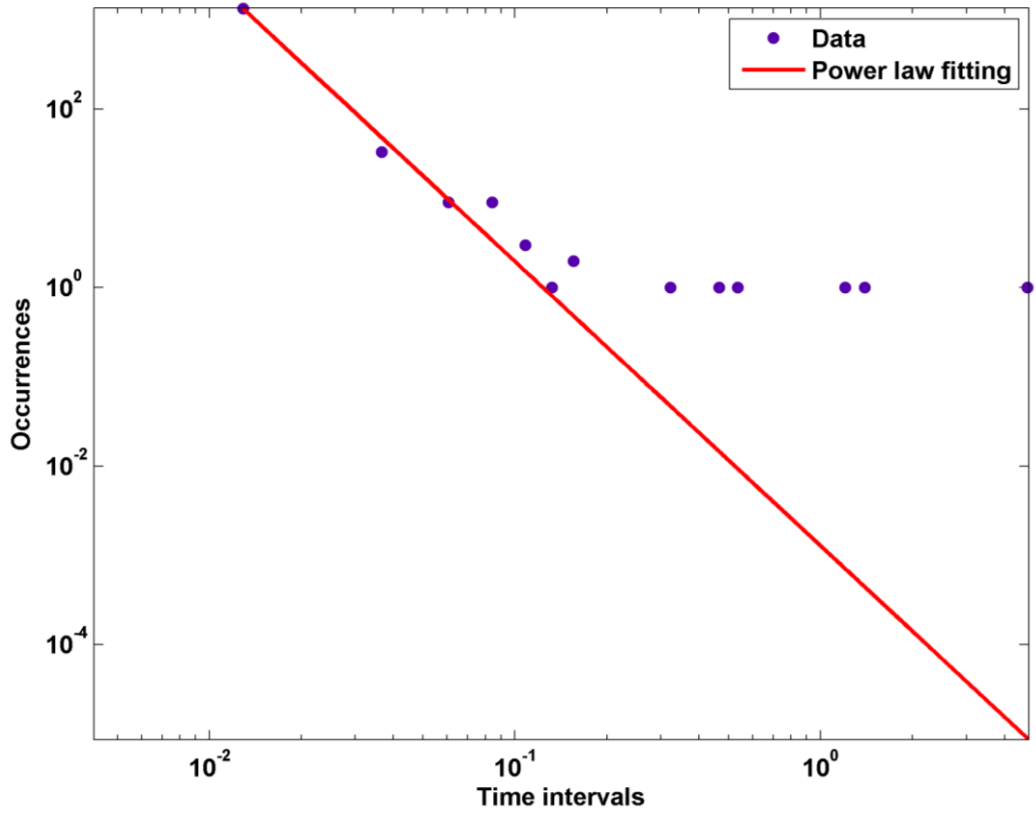

**Fig. 8**: power law fitting of the distribution of subsequent time intervals relative to all triggering phase, up to complete landslide detachment (R-square=1).

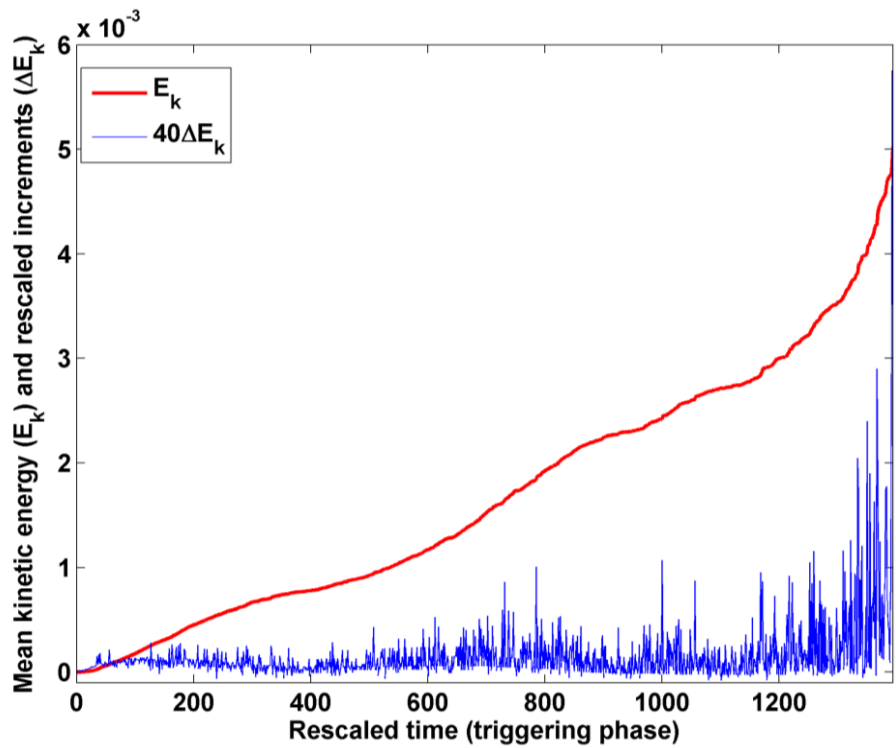

**Fig. 9**: mean kinetic energy (red line) and mean kinetic energy increments (blu line), the latter is rescaled only for visualization.

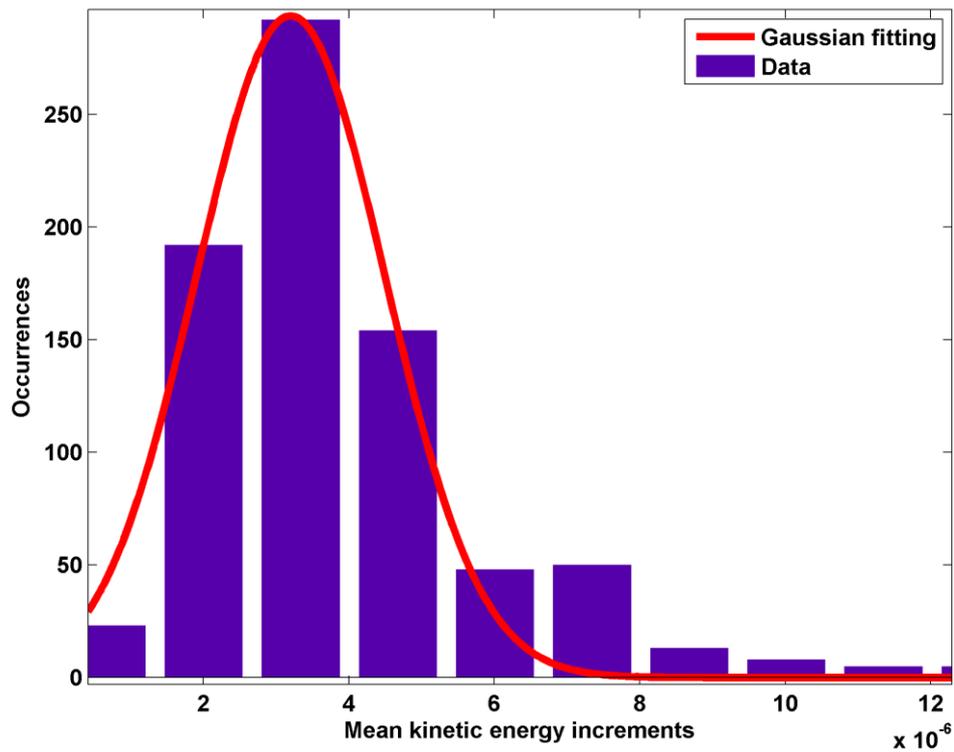

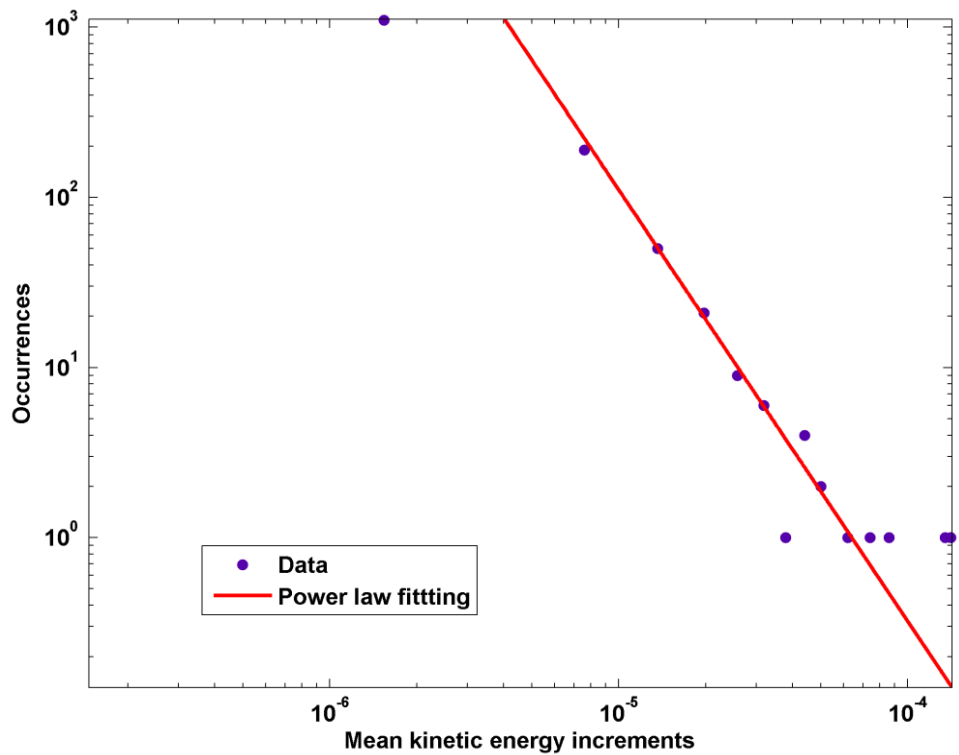

**Fig. 10**: in the top the Gaussian distribution of mean kinetic energy increments relating to the initial phase of triggering (R-square=0.9903), in the bottom the power law distribution relating to all times up to final phase of triggering (R-square=0.9994).

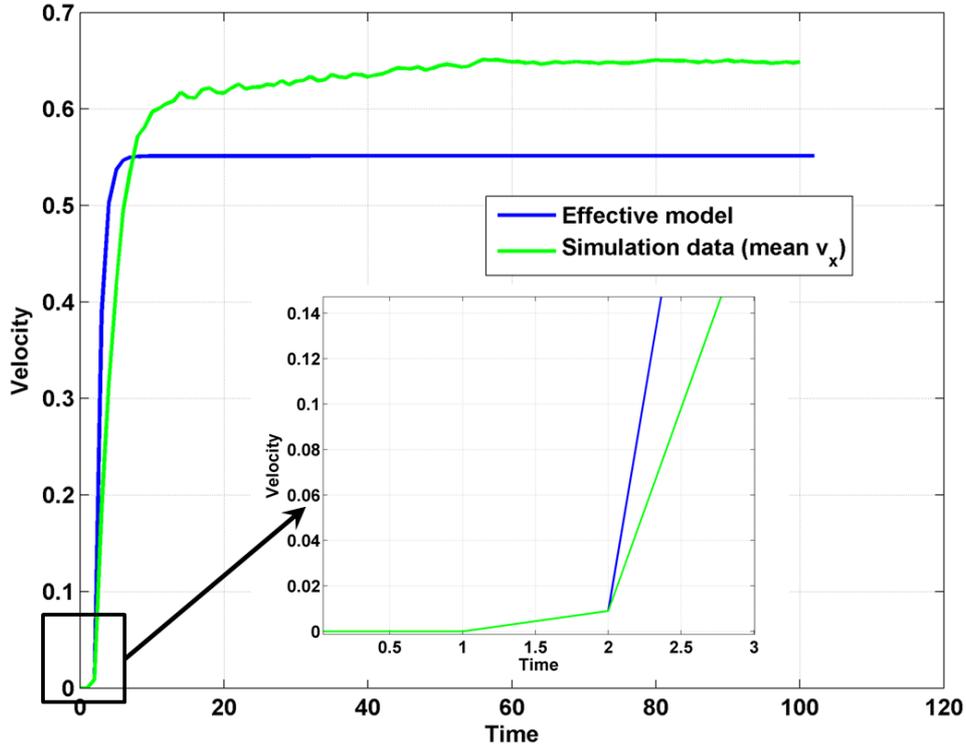

**Fig. 11**: comparison between the mean velocity (component along inclined planed) obtained from simulation (green line) and the velocity of the one-dimensional effective model.

### 3.2 Simulation 2

In the second numerical experiments we consider a system of spherical particles with different radii in a random configuration, arranged in an inclined plane with vertical infiltration, as shown in Figure 1. As in simulation 1 we impose rigid boundary condition and we use the force scheme of eq. (11) and the second order algorithm to update positions and velocities (eq. 15) with $\Delta t = 0.001$. Also in this case we simulate a system of 3000 particles with a duration of $10^5$ temporal steps, i.e., $10^2$ in time). Now we consider the density of the spherical particles constant and equal to $8 \cdot 10^{-3}$ and the radii vary uniformly: with this choice the radii vary in the range $\sim[0.5, 0.6666]$ and the masses in the range $\sim[0.0042, 0.0099]$. In this simulation we adopt the constant values indicated in Table 2.

| | |
|---|---|
| Slope angle ($\varphi$) | 60° |
| Friction angle ($\phi'$) | 40° |
| Fractal order temporal derivative ($\alpha$) | 1 |
| Fractal order space derivative ($\beta$) | 2.5 |
| Fractal diffusion coefficient ($D$) | 100 |
| Repulsive coefficient of interacting force ($k_1$) | 0.003 |
| Attractive coefficient of interacting force ($k_2$) | 0.001 |
| Gravity acceleration ($g$) | 1 |
| Coefficient of cohesion ($c_{k0}$) | 0.5 |
| Upper friction coefficient ($\mu_{d0}$) | 0.69 |
| Lower friction coefficient ($\mu_{dlow0}$) | 0.4 |
| Viscosity coefficient ($\gamma$) | 0.01 |

**Table 2**: values of the model constant for simulation 2.

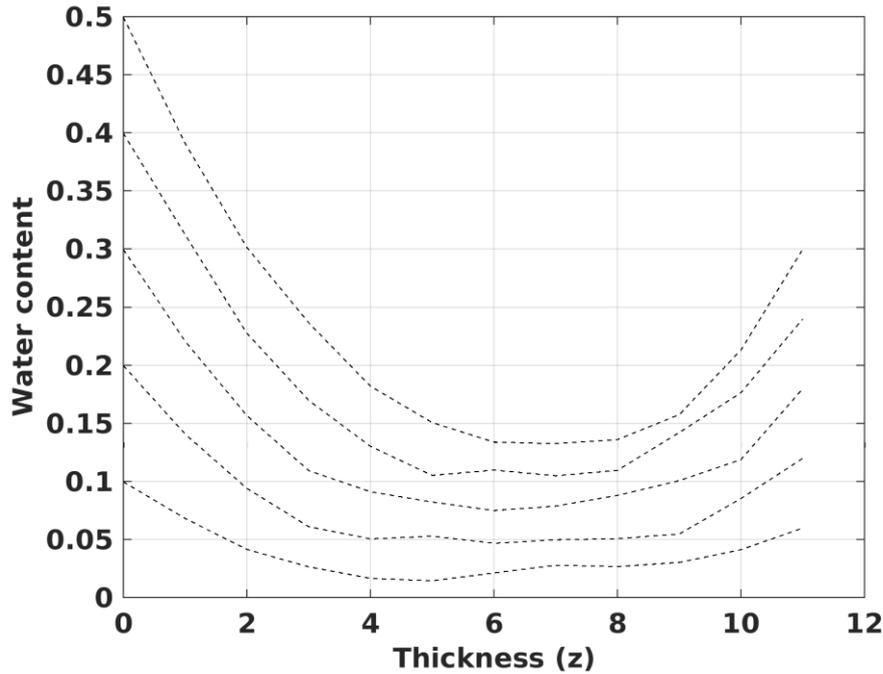

**Fig. 12**: water content in space (thickness of the system) and in time (curve from bottom to top of the figure) for a single vertical layer in simulation 2.

In Figure 12 we report the trend of water content curve in time and space. We adopt non homogeneous boundary conditions ($\theta(0,t)=a\cdot t$, $\theta(L,t)=d\cdot t$, $\theta(z, 0)=0$). In this case we consider the possibility of water accumulation in the soil depth in time according to the condition $\theta(L,t)=d\cdot t$, where $d$ is a constant and $L$ is the thickness of the soil. In Figure 13 we report some configuration of the system at different times where the red particles are at rest and the green ones are in motion. Moreover the right bottom graph of Figure 13 represents the final configuration (the same of bottom left graph) exploiting the real dimension of the slope and masses of the landslide as in simulation 1. In this simulation we observe a segregation effect due to different volume of the particles as we can see in Figure 14. The particles with greater volume tend to segregate to the bottom and to the front of landslide. For the final configuration we study the statistic of the particles with greater radius (range ~[0.62, 0.6666]), i.e., the "blue" particle in Figure 14. As the landslide tends to reduce in thickness we consider 4 horizontal layers and we calculate the number of "blue" particles in each layer. The results is a power law as shown in Figure 14. Then we report the same statistical analysis of simulation 1 regarding the triggering time intervals and the mean kinetic energy increments. The results are similar to simulation 1 (see Figure 15-17). In Figure 18 we show the comparison between the mean velocity (component along inclined planed) obtained from simulation (green line) and the velocity of the one-dimensional effective model. The result is similar, but more accurate then simulation 1 due to the use of a second order algorithm to update positions and velocities of the particles. Analyzing the distribution of each component of velocity, we confirm the results of simulation 1.

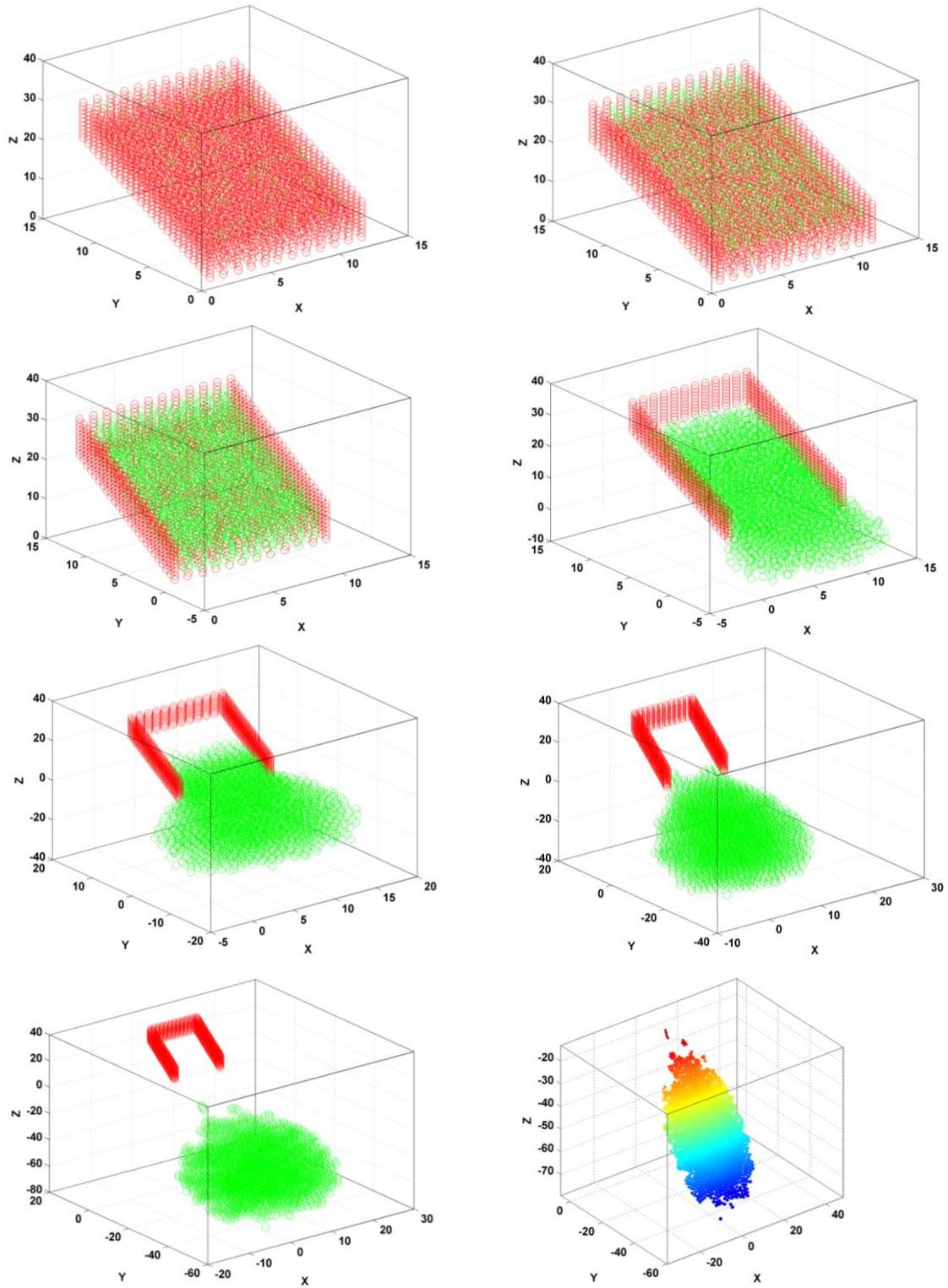

**Fig. 13**: configuration of the landslide at different time starting from left to right and from top to bottom. Red particles are at rest, while green particles are in motion.

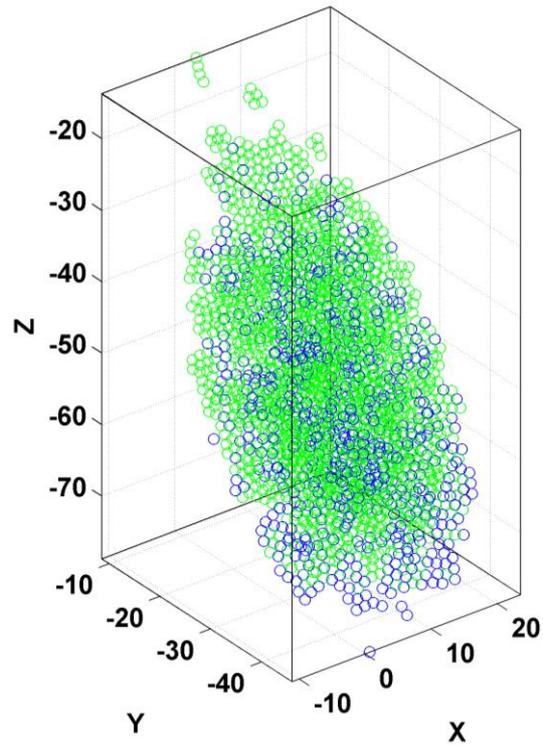

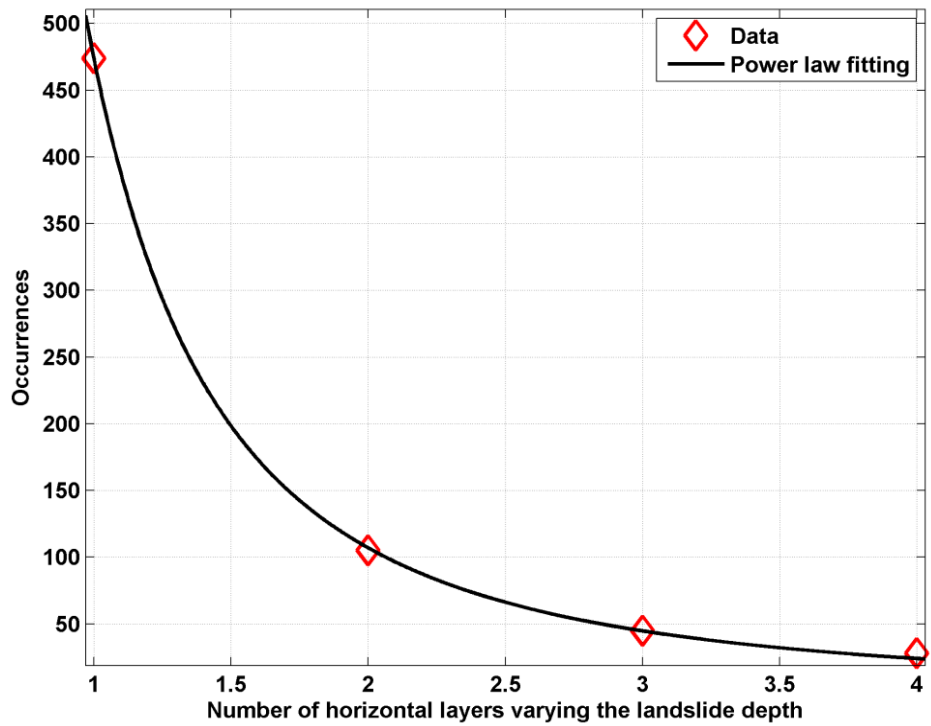

**Fig. 14**: in the top figure we report the final configuration of the landslide where the blue particles have radius in the range ~[0.62, 0.6666] and the green in ~[0.5, 0.62), in the bottom figure we show the power law fitting regarding the statistic of blue particle that segregate downword (R-square=0.9998).

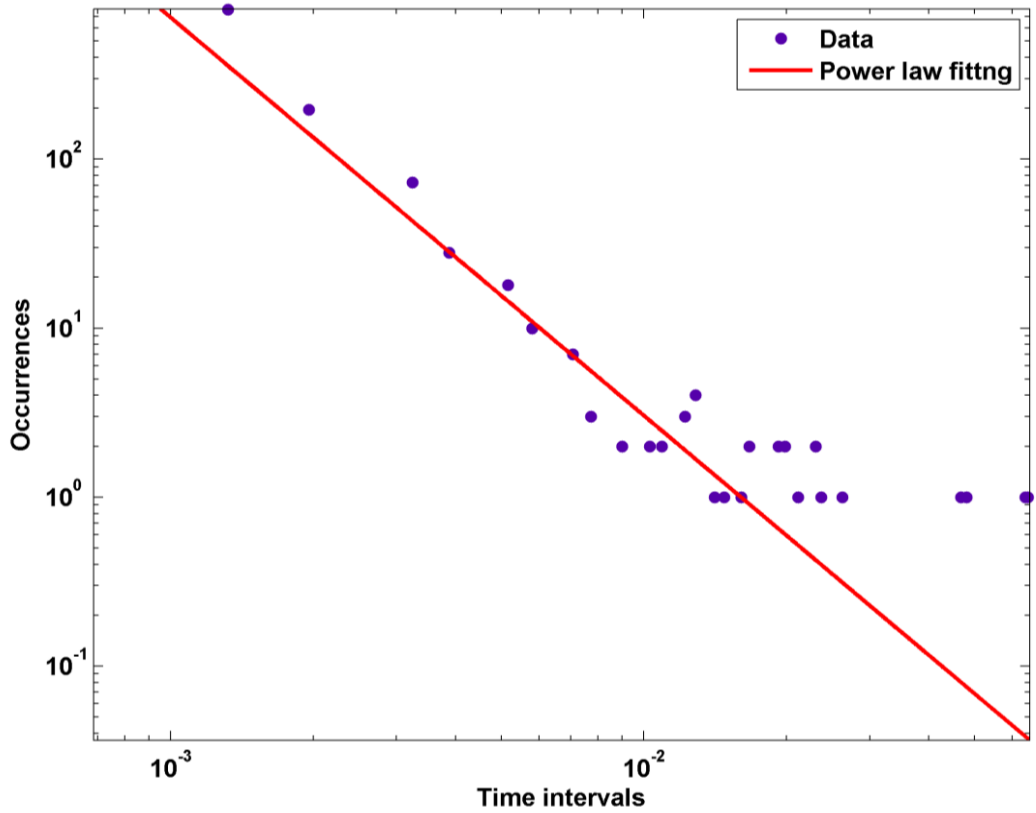

**Fig. 15**: power law fitting of the distribution of subsequent time intervals relative to all triggering phase, up to complete landslide detachment (R-square=0.9996).

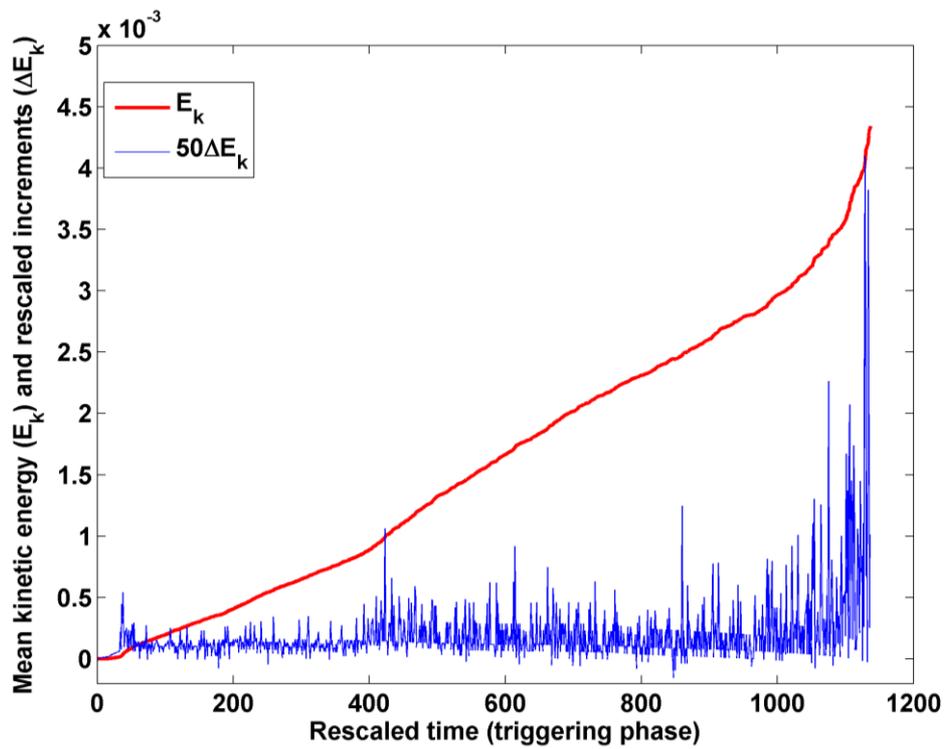

**Fig. 16**: mean kinetic energy (red line) and mean kinetic energy increments (blu line), the latter is rescaled only for visualization.

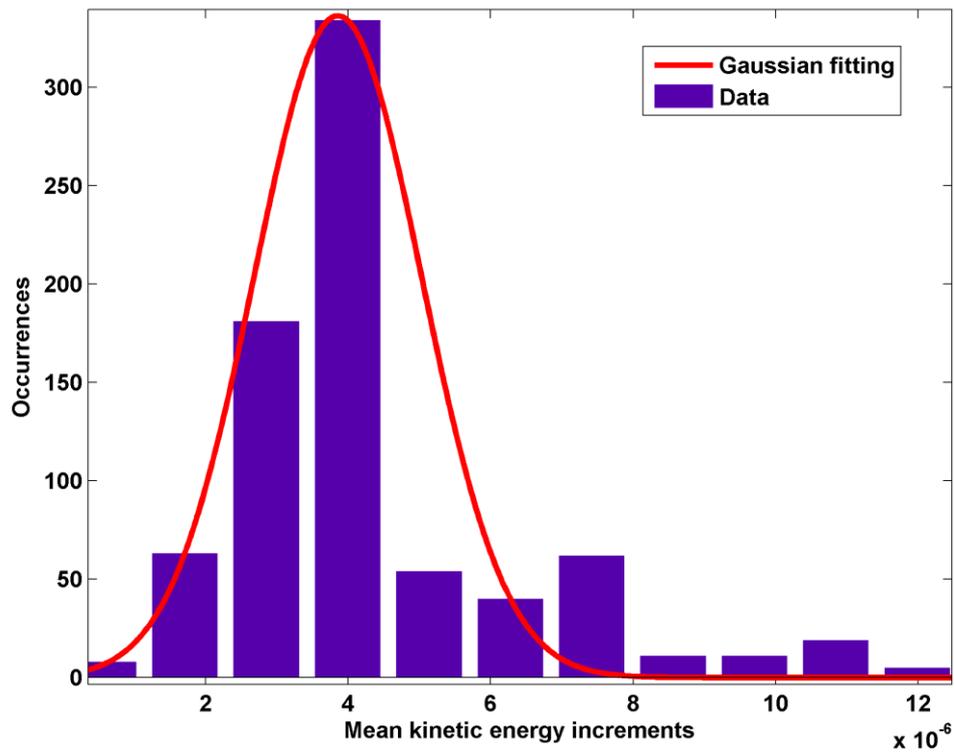
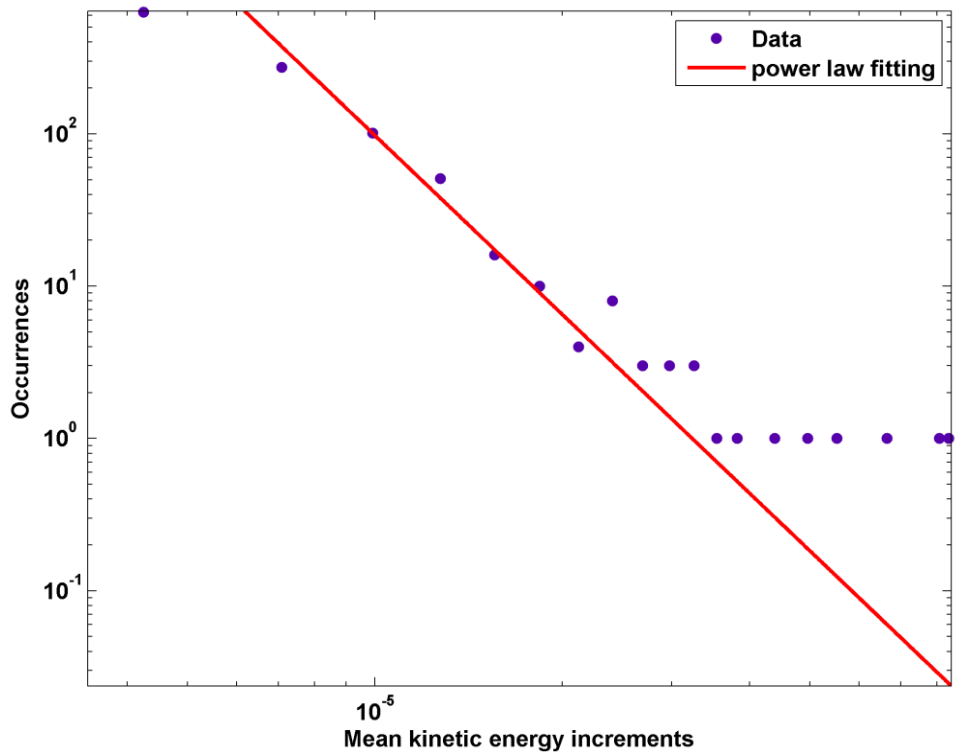

**Fig. 17**: in the top the Gaussian distribution of mean kinetic energy increments relating to the initial phase of triggering (R-square=0.9769), in the bottom the power law distribution relating to all times up to final phase of triggering (R-square=0.9985).

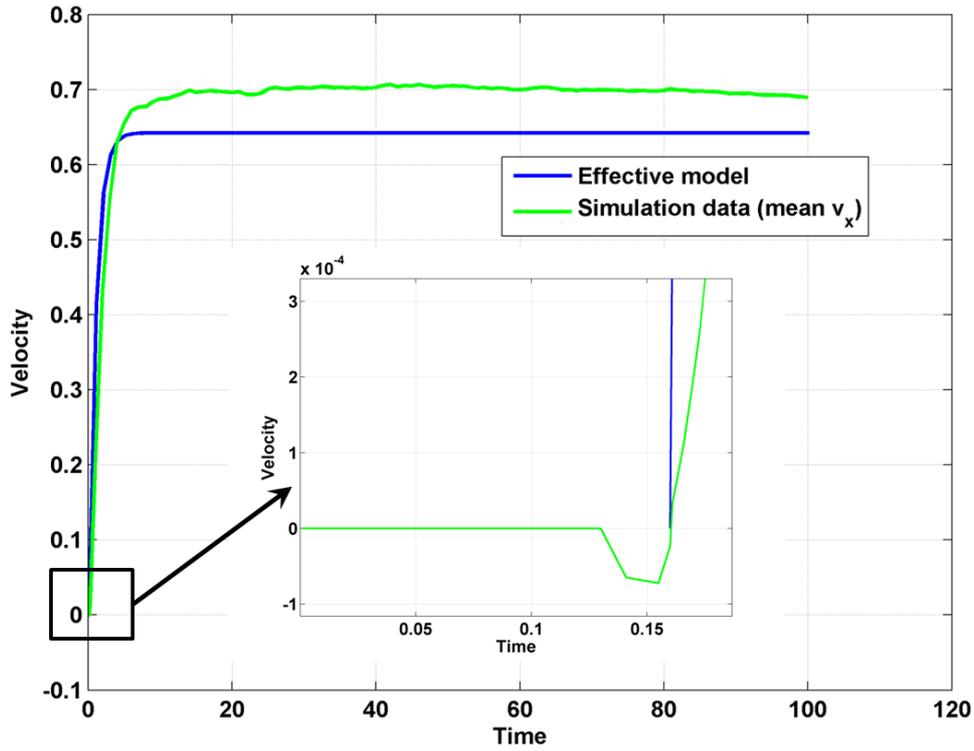

**Fig. 18**: comparison between the mean velocity (component along inclined planed) obtained from simulation (green line) and the velocity of the one-dimensional effective model.

### 3.3 Sensitivity analysis

In this section we report the sensitivity analysis on the main parameters of the model as in the previous work [29,30]. Regarding the simulations, for a better understanding of the parameter influence on the model behavior, we remove all stochastic variations. First of all we study the static constants, i.e., triggering related ones, in particular the order derivatives of fractal diffusion PDE. We do not report the sensibility analysis of some parameters as cohesion, friction or slope angle as this simulations do not produce significant relevance compared to a previous work (see [30]): really, for example, if we study the variation of triggering time as a function of cohesion or slope angle, considering the first particle in motion, this behavior is not influenced by the spatial dimension of the model (2D or 3D). Therefore in Figure 19 and in Figure 20 we show respectively the trend of landslide triggering time varying the time and the space derivative fractal order. Moreover we report, for further information, some curves varying the slope angle. We note that, maintaining constant the derivative spatial order (equal to 2) and decreasing the time derivative fractal order starting from 1 (sub-diffusive behavior), we obtain a lower water content and accordingly a higher triggering time. The triggering time exhibits the same behavior in the sub-diffusive regime varying the spatial derivative fractal order (Figure 20). All obtained curves are fitted by power law. Then we report some trends regarding the parameters that influence the landslide dynamics and propagation. We analyze the landslide final mean velocity (the simulations are stopped after $10^4$ time step) varying the friction coefficients $\mu_{dlow0}$ and maintaining constant $\mu_{d0}$ (equal to 1.4). We consider the lower friction coefficient as the infiltration effects induce a rapid reduction of friction force and therefore $\mu_{dlow0}$ influence the landslide dynamics more than $\mu_{d0}$. In Figure 21 we show the trend (fitted by power law) of the sensibility analysis of the lower friction coefficient. Then in Figure 22 we report the final mean velocity of the simulated landslide versus the viscosity coefficient. Also in this case we obtain a power law curve fitting. Finally in Figure 23 we report the final mean velocity versus the coefficient $k_1$ and $k_2$ of the interacting force. In each simulation we vary a single interacting force coefficient, while the second one is maintained constant. As one might expect, noting that we have small variations in the value of the landslide final mean velocity, the trend is not regular due to the non-linear expression of the interacting force.

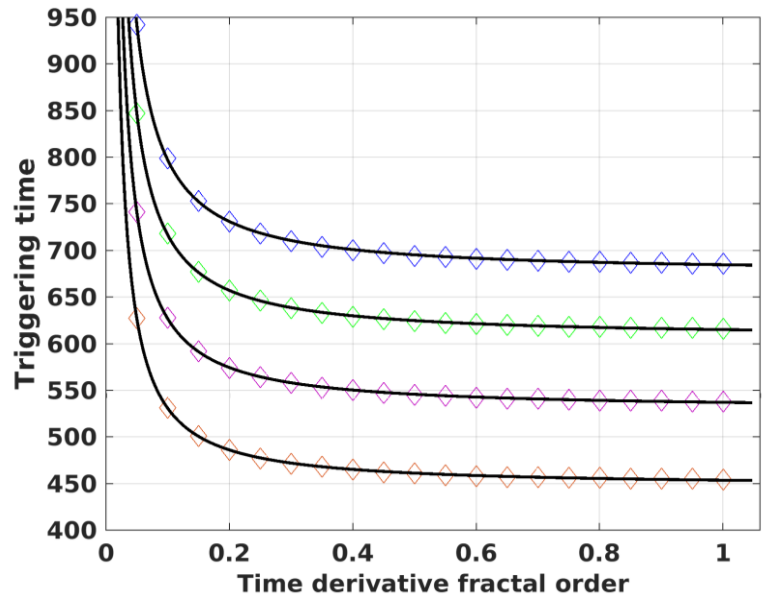

**Fig. 19**: the triggering time varying the time derivative fractal order, we report 4 trends for 4 slope angle (40° blue diamonds, 50° green diamonds, 60° magenta diamonds, 70° red diamonds); the black curves represent the power law fitting (R-square=0.9998).

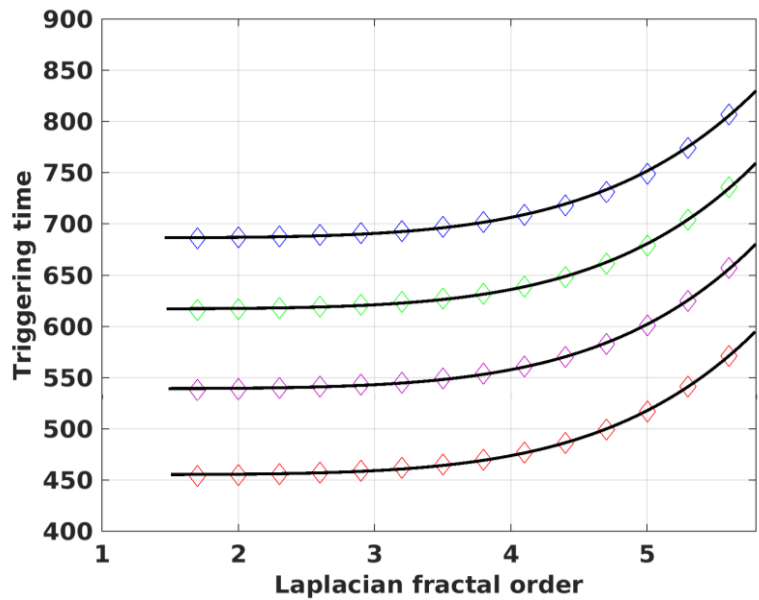

**Fig. 20**: the triggering time varying the spatial derivative fractal order, we report 4 trends for 4 slope angle (40° blue diamonds, 50° green diamonds, 60° magenta diamonds, 70° red diamonds); the black curves represent the power law fitting (R-square=0.9996).

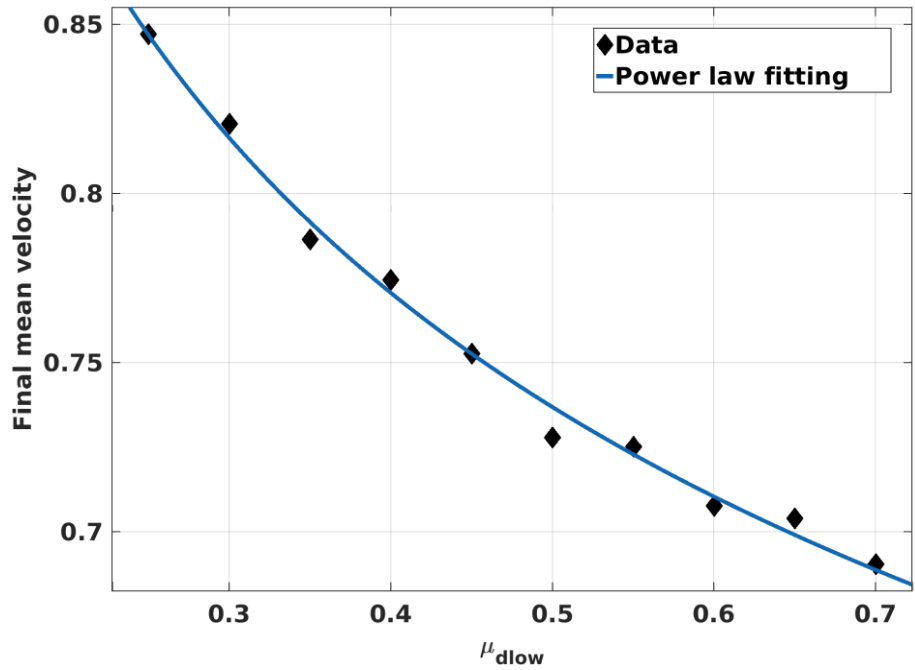

**Fig. 21**: the landslide final mean velocity versus the lower friction coefficient; the black curves represent the power law fitting (R-square=0.9929).

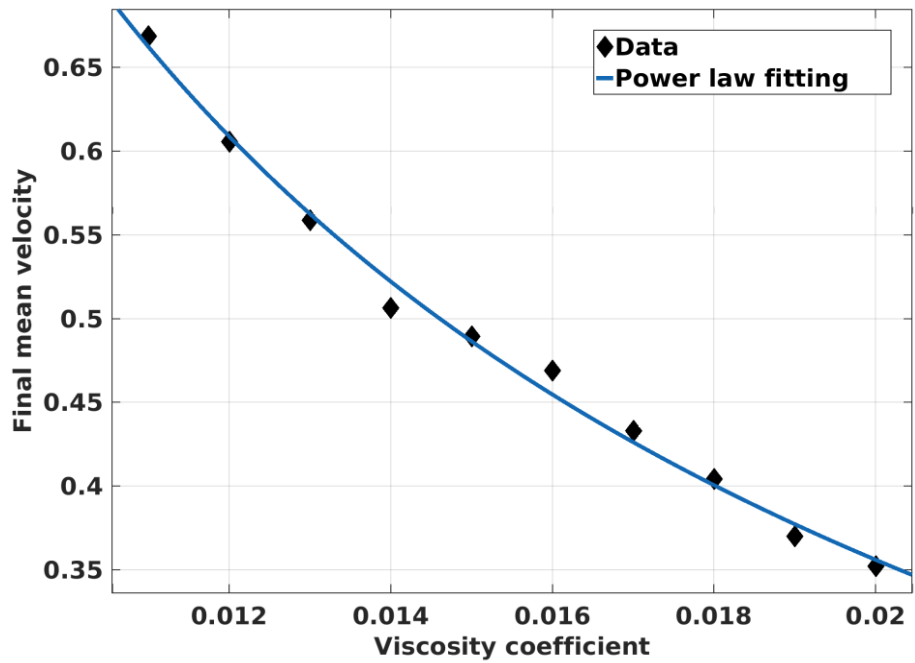

**Fig. 22**: the landslide final mean velocity versus the viscosity coefficient; the black curves represent the power law fitting (R-square=0.983).

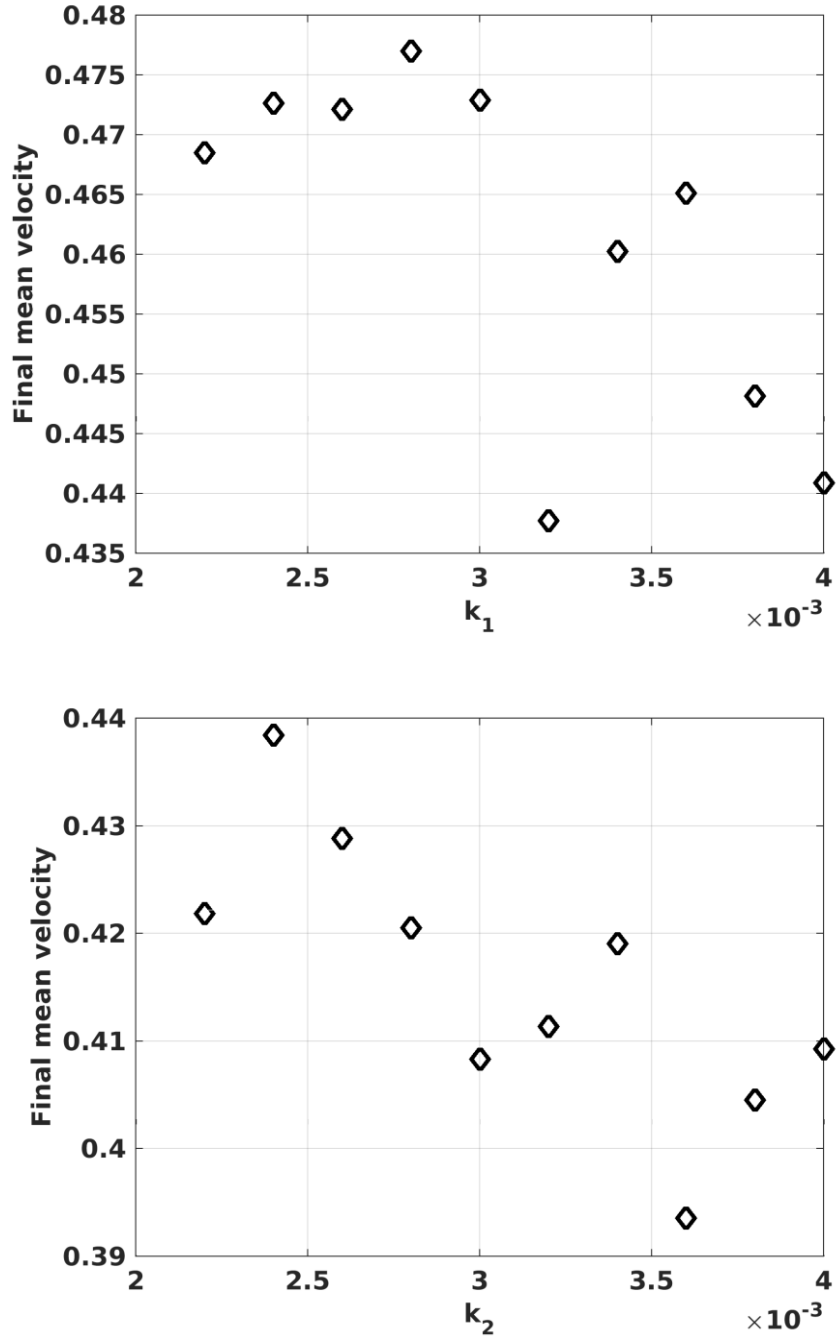

**Fig. 23**: the landslide final mean velocity versus the interacting force coefficients, in the top graph we report the trend of the repulsive coefficient $k_1$ ($k_2 = 0.002$), while in the bottom the attractive one $k_2$ ($k_1 = 0.004$).

## 4 Discussion and conclusions

In this paper we propose a three-dimensional lagrangian model of rain induced landslide in which we consider both triggering and propagation phases. Here we extend previous two-dimensional model illustrated in [29,30]. A 3D scheme is more realistic and in particular permits a better representation of the propagation dynamics. The main originality of the manuscript lies in the use of fractal calculus to take into account the water infiltration processes that perturb the pore pressure, considered in the model as a scalar field. The increasing of the latter represents the main triggering factor of rain induced landslide. Moreover a fractal/fractional approach, as discussed in this paper, is suitable for a correct physical description of the infiltration processes in porous system. We calculate analytically the solution of fractal diffusion PDE and we develop a numerical integration scheme of the latter. Thus this numerical scheme is integrated in the MD algorithm to consider the triggering mechanism. In the model we introduce stochastic variations of the parameters, necessary to consider the variability of the real soils. Therefore we have discussed two different simulations analyzing the behavior of the system from statistical point of view. As shown, some statistical observables exhibit a power law distribution that confirms the complex and extreme nature of this type of phenomena. Indeed also a minimal schematization of a landslide shows a variety of behaviors [47]. It is known that other natural hazards (earthquakes and forest fire) exhibit similar distribution [48,49]. We confirm the results of the previous works [29,30], observing characteristic energy and velocity pattern typical of real landslides [50]. It is possible also to observe fracture, detachment, arching phenomena and zones with higher compression. To demonstrate the coherence of the simulations, we test the component of the mean velocity, along the propagation axis, with a one-dimensional effective model. Then our simulations, using particles of different radius and same density, exhibit segregation effects as shown in Figure 14. Finally we report a sensitivity analysis of main parameters that confirm the consistency of the model and simulations. In conclusion our model can reproduce the behavior of real landslides, it is congruent with granular matter behavior (see for example [51]) and therefore it could be tested for the application to real cases using, eventually, data integrated in a digital terrain model.

## Acknowledgements

We thank the Ente Cassa di Risparmio di Firenze for its support under the contract "*SVILUPPO DI MODELLI MATEMATICI PER L'INNESCO DI FRANE SU SCALA REGIONALE/NAZIONALE (APPLICAZIONI A SCOPO DI PROTEZIONE CIVILE) E DI MODELLI TEORICI DI INNESCO E PROPAGAZIONE SU SCALA DI VERSANTE"*.